\documentclass[aip,jcp,twocolumn,reprint,showpacs,superscriptaddress,floatfix]{revtex4-1}
\usepackage{amstext,amssymb}
\usepackage[fleqn]{amsmath}
\usepackage[dvips]{color}
\usepackage{indentfirst}
\usepackage{makeidx}
\usepackage{slashed}
\usepackage{color}
\usepackage{graphicx}
\newcommand{\sm}{\scriptscriptstyle}               
\newcommand{\avl}{\big\langle}                     
\newcommand{\avr}{\big\rangle}                     
\newcommand{\be}{\beta}                            
\newcommand{\ch}{\mathrm{C}}                       
\newcommand{\dm}{\rho}                             
\newcommand{\dk}{\gamma}                           
\newcommand{\en}{\mathrm{E}}                       
\newcommand{\h}{\hbar}                             
\newcommand{\iti}{t_{\sm{\mathrm{o}}}}             
\newcommand{\id}{\mathbb{I}}                       
\newcommand{\p}{\prime}                            
\newcommand{\drm}{\mathrm{d}}                      
\newcommand{\e}{\mathop{\mathrm{e}}\nolimits}      
\newcommand{\iu}{i\,}                              
\newcommand{\sd}{\mathrm{J}}                       
\newcommand{\sdm}{D}                               
\newcommand{\sdmt}{\breve{D}}                      
\newcommand{\sdmtd}{\bar{D}}                       
\newcommand{\ep}{\lambda}                          
\newcommand{\tf}{\Delta}                           
\newcommand{\tp}{\otimes}                          
\newcommand{\w}{\omega}                            
\newcommand{\Wre}{\tilde{W}^{\prime}}              
\newcommand{\Wim}{\tilde{W}^{\prime\prime}}        
\begin{document}
\title{Generalized Gibbs state with modified Redfield solution: Exact agreement up to second order}

\author{Juzar~Thingna}
\email[]{juzar@nus.edu.sg}
\affiliation{Department of Physics and Center for Computational Science and Engineering, National University of Singapore, Singapore 117542, Republic of Singapore}
\author{Jian-Sheng~Wang}
\affiliation{Department of Physics and Center for Computational Science and Engineering, National University of Singapore, Singapore 117542, Republic of Singapore}
\author{Peter H\"anggi}
\affiliation{Department of Physics and Center for Computational Science and Engineering, National University of Singapore, Singapore 117542, Republic of Singapore}
\affiliation{Institut f\"ur Physik Universit\"at Augsburg, Universit\"atsstrasse 1, D-86135 Augsburg, Germany}

\date{28 March 2012, revised 2 May 2012}
\begin{abstract}
A novel scheme for the steady state solution of the standard Redfield quantum master equation is developed which yields agreement with the exact result for the corresponding reduced density matrix up to second order in the system-bath coupling strength. We achieve this objective by use of an analytic continuation of the off-diagonal matrix elements of the Redfield solution towards its diagonal limit. Notably, our scheme does not require the provision of yet higher order relaxation tensors. Testing this modified method for a heat bath consisting of a collection of harmonic oscillators we assess that the system relaxes towards its correct coupling-dependent, generalized quantum Gibbs state in second order. We numerically compare our formulation for a damped quantum harmonic system with the nonequilibrium Green's function formalism: we find good agreement at low temperatures for coupling strengths that are even larger than expected from the very regime of validity of the second-order Redfield quantum master equation. Yet another advantage of our method is that it markedly reduces the numerical complexity of the problem; thus allowing to study efficiently large-sized \emph{system} Hilbert spaces.
\end{abstract}

\maketitle
\section{Introduction}
\label{sec:1}
The application of canonical statistical mechanics inherently assumes large environments interacting weakly with a few relevant degrees of freedom, then yielding the well-known canonical thermal state. The emergence of the thermal steady state canonical Gibbs density matrix at very weak coupling strength, or the generalized thermal Gibbs state at finite coupling strengths, when starting from quantum dynamical microscopic laws still presents a formidable problem. This objective is known under the label of open system quantum dynamics. A main goal is then to obtain the reduced system dynamics in terms of the reduced density matrix, which typically is approached using a wide variety of approximate quantum master equations.

Formally exact generalized quantum master equations yield the reduced system dynamics either within a time-convolution (time-non-local) form\cite{Nakajima,Zwanzig1} and equivalently also in its time-convolutionless (time-local) form\cite{Fulinski, Tokuyama,Hanggi1,Grabert1,Shibata,Nan}. The hierarchy equations of motion approach\cite{Tanimura,Yan,Shao,Xu} yields yet another, formally exact approach in terms of an infinite number of auxiliary reduced density operators. All these formally exact approaches are computationally very demanding and, typically, can treat systems possessing a small Hilbert space dimension only. For these formally exact approaches it is only for specific setups, such as the situation involving (i) a system of harmonic oscillators\cite{Grabert2,Grabert3,Grabert4,Riseborough,Hu,Chou,Zerbe,Dhar}, (ii) the intricate dissipative Landau-Zener dynamics at zero temperature\cite{Wubs}, or (iii) the known cases with a strictly pure dephasing dynamics\cite{Luczka,vanKampen,Doll}, that the exact solutions can be obtained.

Timely applications, however, call for a definite need to study systems which span a rather large Hilbert space for its underlying nonlinear dynamics. In absence of analytic exact results the quantum master equations are typically evaluated using perturbation theory in the system(S)-bath(B) coupling strength. Commonly, the perturbation is truncated to second order in system-bath coupling, resulting in a whole group of approximate quantum master equations\cite{Pauli,Lindblad,Redfield,Esposito,Blum}. Out of these many existing approximation schemes the Redfield quantum master equation (RQME) is the most generic one from which the Pauli\cite{Pauli} and the Lindblad\cite{Lindblad,Alicki} master equations can be deduced upon invoking further approximations\cite{Breuer}. Sometimes the RQME is also subjected to the secular approximation\cite{Wangsness,Laird,Kohen1,Fleming1} and/or one neglects the Lamb shift-type contributions\cite{Pollard,Kohen2,Geva1}, which cannot always be justified a priori. Generally, all these diverse approximations, even within the weak coupling limit, do fail at zero temperature. This is so because of the neglect of algebraic long-time tail contributions stemming from zero-temperature bath correlations\cite{Hanggi2}. Even without invoking such approximations, a question recently raised is the overall accuracy of the Redfield formalism\cite{Mori,Fleming2}: Therein these authors demonstrated that the Redfield formalism is not correct for the steady state within its commonly used second order form. The discrepancy arises due to the second order diagonal elements which require contributions from the higher order relaxation tensor for their correct evaluation. In view of these findings it is not possible to capture correctly the effects of finite coupling up to second order by use of the Redfield formalism; -- this feature also corrects some inadequately stated claims contained in the previous literature\cite{Geva2,Romero}.

Our main goal with this study is to correctly evaluate the steady state reduced density matrix up to second order in the system-bath coupling \emph{without having to invoke higher order relaxation tensors}. In order to successfully achieve this objective we put forward a modified solution of the Redfield quantum master equation by a procedure that uses the off-diagonal structure in second order to let approach its diagonal structure via a unique analytic continuation.

Comparing this modified Redfield solution with rigorous canonical perturbation theory\cite{Romero,Geva2,Mori}, we show that the modified solution agrees with the exact reduced thermal equilibrium density operator; i.e. the generalized Gibbs state, reading: $\dm =\mathrm{Tr_{\sm{B}}}(\e^{-\be H_{\sm{\mathrm{tot}}}})/\mathrm{Tr}(\e^{-\be H_{\sm{\mathrm{tot}}}}) $, up to second order in the system-bath coupling strength. Our solution is not only accurate as compared to the RQME, but is also numerically efficient. This is because the inherent computational complexity in our method is of O($N^{3}$), where $N$ denotes the dimension of the system Hilbert space. Therefore, our technique enables one to quantum mechanically investigate the small-to-intermediate coupling strength regime for systems possessing a large Hilbert space dimension.

The paper is organized as follows: In Sec.~\ref{sec:2} we describe our basic approach to model quantum dissipation and detail the RQME. In Sec.~\ref{sec:3} we elucidate the insufficient accuracy issue in the second-order steady state Redfield formalism. This part is followed by the exposition of our modified solution to the Redfield quantum master equation. In Sec.~\ref{sec:4} we consider a general nonlinear system that is connected to a harmonic bath and in the long-time limit show that it reaches the generalized Gibbs distribution within canonical perturbation theory carried out up to second order in system-bath coupling. In Sec.~\ref{sec:5} we present the numerical comparison between our modified Redfield solution with the exact solution for a damped harmonic oscillator. We find a considerable improvement between exact results and modified solution over \emph{extended} regimes of weak-to-intermediate system-bath coupling strengths for which both the Redfield solution and the Lindblad solution fail. Sec.~\ref{sec:6} summarizes our main findings while the Appendix details canonical perturbation theory.
\section{Redfield Quantum Master Equation in presence of archetype quantum dissipation}
\label{sec:2}
The basic approach to model quantum dissipation has been studied extensively before. The model Hamiltonian for the bath and system-bath coupling has a long-standing history\cite{Hanggi2,Bogolybov,Magalinskii,Sentizky,Ford} but goes under the label of Zwanzig-Caledira-Leggett model\cite{Zwanzig,Caldeira1,Caldeira2, Weiss},
\begin{eqnarray}
\label{eq:no1}
H_{\sm{\mathrm{tot}}}&=&H_{\sm{\mathrm{S}}}+H_{\sm{\mathrm{B}}}+H_{\sm{\mathrm{RN}}}+H_{\sm{\mathrm{SB}}},
\end{eqnarray}
where
\begin{eqnarray}
\label{eq:no2}
H_{\sm{\mathrm{S}}}&=&\frac{p^{2}}{2M} +V(q)
\end{eqnarray}
denotes the generally nonlinear system Hamiltonian of a particle of mass $M$ moving in a potential $V(q)$. Here
\begin{eqnarray}
\label{eq:no3}
H_{\sm{\mathrm{B}}}&=&\sum_{n=1}^{\infty}\left(\frac{p_{n}^{2}}{2m_{n}}+\frac{m_{n}\w_{n}^{2}}{2}x_{n}^{2}\right),
\end{eqnarray}
describes the thermal environment as an infinite collection of harmonic oscillators, each having a mass $m_{n}$ and a frequency $\w_{n}$.
\begin{eqnarray}
\label{eq:no4}
H_{\sm{\mathrm{RN}}}&=&S^{2}\left(\frac{1}{2}\sum_{n=1}^{\infty}\frac{c_{n}^{2}}{m_{n}\w_{n}^{2}}\right),
\end{eqnarray}
is the potential renormalization in which the variable $S$ denotes any function of the system variables $p$ and $q$ and
\begin{eqnarray}
\label{eq:no5}
H_{\sm{\mathrm{SB}}}&=&S\tp B \nonumber \\ 
&=& S\tp \left(-\sum_{n=1}^{\infty} c_{n} x_{n}\right), 
\end{eqnarray}
is the system-bath coupling Hamiltonian, wherein the $c_{n}$ denotes the system-bath coupling constant of the $n$-th oscillator with the system operator $S$. The collective bath operator is $B = -\sum_{n=1}^{\infty}c_{n}x_{n}$. Throughout this work we use $\h=1$ and $k_{\sm{\mathrm{B}}}=1$.

Using correlation-free initial conditions, i.e., $\dm_{\sm{\mathrm{tot}}}(\iti) = \dm_{\mathrm{\sm{S}}}(\iti)\tp\dm_{\mathrm{\sm{B}}}(\iti)$, with $\dm_{\mathrm{\sm{B}}}(\iti)$ being the canonical thermal state of the bath, and assuming overall weak system-bath coupling, we obtain the perturbative, $2$-nd order Redfield quantum master equation in its time-local energy representation as\cite{Redfield,Weiss,Breuer,Blum},
\begin{eqnarray}
\label{eq:no6}
\frac{\drm\dm_{nm}}{\drm t} &=&-\iu \tf_{nm}\dm_{nm}+\sum_{ij}\mathrm{R}_{nm}^{ij,(2)}\dm_{ij}, \nonumber \\
\mathrm{R}_{nm}^{ij,(2)}&=&S_{ni}S_{jm}\biggl(W_{ni}+W_{mj}^{*}\biggr)-\delta_{j,m} \sum_{l} S_{nl}S_{li}W_{li} \nonumber \\
&&-\delta_{n,i} \sum_{l} S_{jl}S_{lm}W_{lj}^{*},
\end{eqnarray}
where the explicit time dependence in the reduced density matrix $\dm (t) = \mathrm{Tr}_{\sm{\mathrm{B}}}\left(\dm_{\sm{\mathrm{tot}}}(t)\right)$ has been suppressed, i.e., $\dm_{nm}= \avl n|\dm(t)|m\avr$; ${|n\avr}$ being the energy eigenvector of the bare system. Here, the matrix elements $S_{ik}$ are defined as $S_{ik} = \avl i|S|k\avr $. Despite the apparent time-local form of Eq.~(\ref{eq:no6}) the non-Markovian behavior is fully captured due to the time dependence in the transition rates $\tilde{W}$. These are given by,
\begin{eqnarray}
\label{eq:no7}
W_{jk}&=&W_{jk}^{\p}+\iu W_{jk}^{\p\p} \nonumber \\
&=&\Wre_{jk}+\iu \left(\frac{\dk_{\sm{0}}}{2}+\Wim_{jk}\right), \\
\label{eq:no8}
\tilde{W}_{jk}&=&\Wre_{jk}+\iu \Wim_{jk}, \nonumber \\
&=&\int_{0}^{t-\iti} \drm\tau \e^{-\iu \tf_{jk} \tau} \ch(\tau),
\end{eqnarray}
where $\Wre_{jk} = W_{jk}^{\p}$ and $\tf_{jk} = E_{j} - E_{k}$ is the energy difference between system energy levels. Above $\dk_{\sm{0}}/2$ which arises from $H_{\sm{\mathrm{RN}}}$ has been neglected from the uncoupled propagation (Eq.~(\ref
{eq:no8})) and accounted in $W$.

Instead of specifying all the parameters of the bath we now define the spectral density $\sd(\w)$ as,
\begin{eqnarray}
\label{eq:no9}
\sd(\w)&=&\pi\sum_{n=1}^{\infty}\frac{c_{n}^{2}}{2m_{n}\w_{n}}\delta(\w-\w_{n}).
\end{eqnarray}
Using the spectral density we can calculate the damping kernel at time $t=0$; i.e., $\dk_{\sm{0}}$, used in Eq.~(\ref{eq:no7}) as,
\begin{eqnarray}
\label{eq:no10}
\dk_{\sm{0}} & = & \frac{1}{M}\sum_{n=1}^{\infty}\frac{c_{n}^{2}}{m_{n}\w_{n}^{2}} = \frac{2}{M}\int_{0}^{\infty} \frac{\drm\w}{\pi}\frac{\sd(\w)}{\w},
\end{eqnarray}
and the equilibrium bath-bath correlation function $\ch(\tau)=\avl \tilde{B}(\tau)B \avr$, where $\tilde{B}(\tau)$ is evolving according to $\mathrm{exp}\left(-\iu H_{\sm{\mathrm{B}}}\tau\right)$, used in Eq.~(\ref{eq:no8}) as,
\begin{eqnarray}
\label{eq:no11}
\ch(\tau)&=&\int_{0}^{\infty}\frac{\drm\w}{\pi} \sd(\w) \Biggl[\mathrm{coth}\left(\frac{\be\w}{2}\right)\mathrm{cos}\left(\w\tau\right)\Biggr.\nonumber \\
&&\Biggl.-\iu \mathrm{sin}\left(\w\tau\right)\biggr].
\end{eqnarray}
Although the bath-bath correlation defined above is specific to a harmonic bath, the theory presented here can readily be generalized to other bath models, e.g. spin baths as long as the bath-bath correlation $\ch(\tau)$ can be evaluated.
\section{Modified solution to the Redfield quantum master equation}
\label{sec:3}
\subsection{Perturbative accuracy of steady state Redfield solution}
\label{subsec:3.1}
Recently Mori and Miyashita\cite{Mori} and, as well, Fleming and Cummings\cite{Fleming2} independently established that for a generic ``$2n$''-order quantum master equation the solution in the long-time limit can be correct only up to order ``$2n-2$'', because the diagonal elements loose their accuracy over evolving time. These authors suggest that in order to obtain a steady state solution correct up to $2$-nd order a $4$-th order master equation\cite{Jang,Laird,Schroder} should be used, which can be numerically accomplished for small system Hilbert spaces only.

We first corroborate this finding with a different method, concentrating on the steady state accuracy of a $2$-nd order RQME. We start out with the generic perturbation series expansion to all orders in the system-bath coupling of the time-local, formally exact master equation; i.e.,
\begin{equation}
\label{eq:no12}
\frac{\partial\dm}{\partial t} = \left(\bar{\tf}+\sum_{n=2,4,6,\cdots}^{\infty} \ep^{n} \mathrm{R^{(n)}}(t-\iti)\right)\dm,
\end{equation}
and the reduced density matrix,
\begin{eqnarray}
\label{eq:no13}
\dm &=& \sum_{n=0,2,4,\cdots}^{\infty} \ep^{n} \dm^{(n)},
\end{eqnarray}
where $\ep$ is a dimensionless parameter whose power indicates the corresponding order of the perturbation expansion. Eventually, $\ep$ will be set to $1$. $\bar{\tf}$ above is a four tensor depending on the system Hamiltonian. The operator $\mathrm{R^{(n)}}(t-\iti)$ denotes the Redfield superoperator of rank $4$ which depends both on the system operator and the bath correlators. Next we rearrange $\dm$ into a column vector and split it into its diagonal part ($\dm_{\mathrm{\sm{d}}}$) and off-diagonal part ($\dm_{\mathrm{\sm{od}}}$). Then, using the RQME (Eq.~(\ref{eq:no6})) the $0$-th order tensor in Eq.~(\ref{eq:no12}) can be rewritten as a matrix assuming the form,
\begin{eqnarray}
\label{eq:no14}
\bar{\tf} & \equiv & \left( \begin{array}{cc}
                0 & 0 \\
		0 & \bar{\tf}_{22}\end{array} \right),
\end{eqnarray}
where $\bar{\tf}_{22}$ is a diagonal matrix with $\tf_{ij}~(i\ne j)$ forming the diagonal. The four tensors $\mathrm{R^{(n)}}(t-\iti)$ are also split accordingly; i.e.,
\begin{eqnarray}
\label{eq:no15}
\mathrm{R^{(n)}}(t-\iti) & \equiv & \left( \begin{array}{cc}
                                 R_{11}^{(n)}(t-\iti) & R_{12}^{(n)}(t-\iti) \\
		                 R_{21}^{(n)}(t-\iti) & R_{22}^{(n)}(t-\iti)\end{array} \right),
\end{eqnarray}
with no restrictions made for the form of the sub-matrices. For the specific case of $n=2$, $\mathrm{R^{(2)}}(t-\iti)$ is the same as the Redfield tensor given in Eq.~(\ref{eq:no6}).

In order to obtain the steady state we set $\partial\dm/\partial t = 0$ and take the limits $(t-\iti)\rightarrow\infty$. Because the stationary problem is not dependent on time we will drop the parentheses from the tensor, i.e., $\mathrm{R^{(n)}}(\infty) \equiv \mathrm{R^{(n)}}$. Therefore, using Eq.~(\ref{eq:no12}) and Eq.~(\ref{eq:no13}) we obtain:
\begin{equation}
\label{eq:no16}
\left(\bar{\tf}+\sum_{n=2,4,6,\cdots}^{\infty} \ep^n \mathrm{R^{(n)}}\right)\sum_{m=0,2,4,\cdots}^{\infty} \ep^{m} \dm^{(m)} = 0.
\end{equation}

In order to obtain $\dm$ correct up to $2$-nd order we equate the coefficients of the different powers of $\ep$ to zero so that we obtain independent equations to calculate $\dm^{(0)}$ and $\dm^{(2)}$. This implies,
\begin{enumerate}
\item Setting the co-efficient of $\ep^{0}$ equal to zero yields,
\begin{eqnarray}
\label{eq:no17}
\dm_{\mathrm{\sm{od}}}^{(0)} & = & 0.
\end{eqnarray}
\item Setting the co-efficient of $\ep^{2}$ equal to zero implies,
\begin{eqnarray}
\label{eq:no18}
R_{11}^{(2)}\dm_{\mathrm{\sm{d}}}^{(0)} & = & 0,\\
\label{eq:no19}
\bar{\tf}_{22}\dm_{\mathrm{\sm{od}}}^{(2)} & = & -R_{21}^{(2)}\dm_{\mathrm{\sm{d}}}^{(0)}.
\end{eqnarray}
\item Setting the co-efficient of $\ep^{4}$ equal to zero provides the condition,
\begin{eqnarray}
\label{eq:no20}
R_{11}^{(2)}\dm_{\mathrm{\sm{d}}}^{(2)} & = & -R_{12}^{(2)}\dm_{\mathrm{\sm{od}}}^{(2)}-R_{11}^{(4)}\dm_{\mathrm{\sm{d}}}^{(0)}.
\end{eqnarray}
\end{enumerate}
Equation~(\ref{eq:no19}) shows that in order to obtain the $2$-nd order off-diagonal elements we need only the $0$-th order and $2$-nd order relaxation tensors which can be obtained from the RQME using Eq.~(\ref{eq:no6}). In contrast, in order to obtain the $2$-nd order diagonal elements from Eq.~(\ref{eq:no20}) one requires knowledge of the $4$-th order relaxation tensor $R_{11}^{(4)}$.
\subsection{Analytic continuation procedure for diagonal density matrix elements}
\label{subsec:3.2}
In this section we present the procedure to obtain the stationary reduced density matrix that is correct up to $2$-nd order in the system-bath coupling without the need to invoke the use of the $4$-th order relaxation tensor. The $0$-th order and the $2$-nd order off-diagonal elements can be obtained correctly from the RQME as described above. Therefore, we use Eq.~(\ref{eq:no17}) and Eq.~(\ref{eq:no18}) along with the Redfield tensor $\mathrm{R}^{(2)}$ detailed in Eq.~(\ref{eq:no6}) to arrive at the $0$-th order reduced density matrix,
\begin{eqnarray}
\label{eq:no21}
\sum_{i}\left(S_{ni}S_{in}\Wre_{ni}-\delta_{n,i}\sum_{l}S_{nl}S_{li}\Wre_{li}\right)\dm_{ii}^{(0)} &=& 0,\\
\dm_{ij}^{(0)} & = & 0, ~~~~(i \neq j). \nonumber
\end{eqnarray}

The $2$-nd order off-diagonal elements follow from Eq.~(\ref{eq:no19}) as,
\begin{eqnarray}
\label{eq:no22}
\dm_{nm}^{(2)} & = & \frac{1}{\iu\tf_{nm}} \sum_{i} S_{ni}S_{im} \biggl[ \Bigl(W_{ni}+W_{mi}^{*}\Bigr)\dm_{ii}^{(0)} \biggr. \nonumber\\
& & \biggl.-W_{in}^{*}\dm_{nn}^{(0)}-W_{im}\dm_{mm}^{(0)}\biggr], ~~~~(n \neq m).
\end{eqnarray}

Note that if we next construct the diagonal elements by merely substituting $n=m$ in Eq.~(\ref{eq:no22}), then the equation exhibits an indeterminate $0/0$ singularity. This indicates that even though we cannot substitute $n=m$ directly, the limit $m \rightarrow n$ might exist. If such a limit indeed exists and being unique, then by use of the uniqueness theorem the $2$-nd order diagonal elements can be obtained by this limiting procedure. In order to perform this limit $m\rightarrow n$ we consider each element of the $2$-nd order reduced density matrix to be a function of the bare system energies $E_{i}$ ($i=1,\cdots, N$). In the energy parameter space we vary only one of the energies $E_{m}$ and let it continuously approach the energy $E_{n}$, via a small complex parameter $z$; i.e., we set $E_{m} \rightarrow E_{n}-z$.

In doing so, we start by splitting the transition rates in Eq.~(\ref{eq:no22}) into its real and its imaginary parts, using Eq.~(\ref{eq:no7}) to obtain:
\begin{eqnarray}
\label{eq:no23}
\dm_{nm}^{(2)} & = & \frac{1}{\iu\tf_{nm}} \sum_{i} S_{ni}S_{im} \Biggl[ \biggl(\Wre_{ni}+\Wre_{mi}\biggr)\dm_{ii}^{(0)} \Biggr. \nonumber\\
& & \Biggl.-\Wre_{in}\dm_{nn}^{(0)}-\Wre_{im}\dm_{mm}^{(0)}\Biggr]\nonumber\\
& &+\frac{1}{\tf_{nm}} \sum_{i} S_{ni}S_{im} \Biggl[ \biggl(\Wim_{ni}-\Wim_{mi}\biggr)\dm_{ii}^{(0)} \Biggr. \nonumber\\
& & \Biggl.+\biggl(\Wim_{in}+\frac{\dk_{\sm{0}}}{2}\biggr)\dm_{nn}^{(0)}-\biggl(\Wim_{im}+\frac{\dk_{\sm{0}}}{2}\biggr)\dm_{mm}^{(0)}\Biggr].
\end{eqnarray}
We next let $E_{m}\rightarrow E_{n}-z$ and perform the limit $z \rightarrow 0$. Therefore, Eq.~(\ref{eq:no23}) becomes,
\begin{eqnarray}
\label{eq:no24}
\dm_{nn}^{(2)} & = & \lim_{z \to 0}\Biggl\{\frac{1}{\iu z} \sum_{i} S_{ni}S_{in} \biggl[ \left(\Wre_{ni}(0)+\Wre_{ni}(-z)\right)\dm_{ii}^{(0)} \biggr.\Biggr. \nonumber\\
& & \Biggl.\biggl.-\left(\Wre_{in}(0)+\Wre_{in}(z)\right)\dm_{nn}^{(0)}\biggr]\Biggr.\nonumber\\
& &\Biggl.+\frac{1}{z} \sum_{i} S_{ni}S_{in} \biggl[ \left(\Wim_{ni}(0)-\Wim_{ni}(-z)\right)\dm_{ii}^{(0)} \biggr.\Biggr. \nonumber\\
& & \Biggl.\biggl.-\left(\Wim_{in}(0)-\Wim_{in}(-z)\right)\dm_{nn}^{(0)}\biggr.\Biggr. \nonumber \\
& & \Biggl.\biggl.+\left(\Wim_{in}(-z)+\frac{\dk_{\sm{0}}}{2}\right)z\frac{\partial\dm_{nn}^{(0)}}{\partial E_{n}}\biggr]\Biggr\},
\end{eqnarray}
where,
\begin{eqnarray}
\label{eq:no25}
\tilde{W}_{ij}(z)&=&\int_{0}^{\infty}\drm\tau\e^{-\iu(\tf_{ij}+z)\tau}\ch(\tau), \nonumber \\
\tilde{W}_{ij}^{*}(z)&=&\int_{0}^{\infty}\drm\tau\e^{\iu(\tf_{ij}+z^{*})\tau}\ch^{*}(\tau).
\end{eqnarray}
Because $\dm_{mm}^{(0)}$ (being the un-normalized $0$-th order reduced density matrix) depends on the energy $E_{m}$ we made use of the Taylor expansion of $\dm_{mm}^{(0)}$ around the energy $E_{n}$ to retain up to the first order:
\begin{eqnarray}
\label{eq:no26}
\lim_{E_{m} \to E_{n}}\dm_{mm}^{(0)}&\simeq&\dm_{nn}^{(0)}+z\frac{\partial\dm_{nn}^{(0)}}{\partial E_{n}}.
\end{eqnarray}
We next define,
\begin{eqnarray}
\label{eq:no27}
V_{ni} & = & \frac{\partial\Wim_{ni}}{\partial\tf_{ni}} = \lim_{z\to 0}\frac{\Wim(0)-\Wim(-z)}{z},
\end{eqnarray}
and note that $\lim_{z\to 0}\Wim_{in}(-z)=\Wim_{in}(0)=\Wim_{in}$. Eq.~(\ref{eq:no24}) can thus be recast as,
\begin{eqnarray}
\label{eq:no28}
\dm_{nn}^{(2)} & = & \sum_{i} S_{ni}S_{in} \left[ V_{ni}\dm_{ii}^{(0)}-V_{in}\dm_{nn}^{(0)}\right]+W_{in}^{\p\p}\frac{\partial\dm_{nn}^{(0)}}{\partial E_{n}}+\bar{\dm}_{nn}^{(2)},\nonumber\\
\end{eqnarray}
where,
\begin{eqnarray}
\label{eq:no29}
\bar{\dm}_{nn}^{(2)} & = &\lim_{z\to 0}\frac{1}{\iu z}\Biggl\{\sum_{i}S_{ni}S_{in}\biggl[\left(\Wre_{ni}(0)+\Wre_{ni}(-z)\right)\dm_{ii}^{(0)} \biggr.\Biggr. \nonumber\\
& & \Biggl.\biggl.-\left(\Wre_{in}(0)+\Wre_{in}(z)\right)\dm_{nn}^{(0)}\biggr]\Biggr\}.
\end{eqnarray}
In the limit $z\rightarrow 0$ it follows from Eq.~(\ref{eq:no25}) that $\lim_{z\to 0}\Wre_{ni}(-z)=\lim_{z\to 0}\Wre_{ni}(z)=\Wre_{ni}(0)=\Wre_{ni}$. Therefore, in this limit the term in the curly bracket in Eq.~(\ref{eq:no29}) assumes precisely the same form as Eq.~(\ref{eq:no21}), hence it is equal to zero. Consequently, Eq.~(\ref{eq:no28}) becomes,
\begin{equation}
\label{eq:no30}
\dm_{nn}^{(2)} = \sum_{i} S_{ni}S_{in} \left[ V_{ni}\dm_{ii}^{(0)}-V_{in}\dm_{nn}^{(0)}+W_{in}^{\p\p}\frac{\partial\dm_{nn}^{(0)}}{\partial E_{n}}\right].
\end{equation}
Eq.~(\ref{eq:no30}) is independent of the way in which the energy $E_{m}$ approaches $E_{n}$ and hence this limit procedure is unique. The uniqueness of the limit is crucial to ensure that the resulting thermal steady state of the system is unique.

The diagonal elements of the density matrix obey the normalization condition $\mathrm{Tr}(\dm) = 1$. Since we performed an analytic continuation to obtain the $2$-nd order diagonal elements there is no guarantee the normalization condition is preserved. Therefore we can write the normalization condition explicitly as,
\begin{eqnarray}
\label{eq:no31}
\dm_{nn}&=&\frac{\dm_{nn}^{(0)}+\dm_{nn}^{(2)}}{\sum_{i}(\dm_{ii}^{(0)}+\dm_{ii}^{(2)})} \nonumber \\
&\simeq&\dm_{nn}^{(0)}+\dm_{nn}^{(2)}-\dm_{nn}^{(0)}\sum_{i}\dm_{ii}^{(2)},
\end{eqnarray}
where we have ignored the $4$-th and higher order terms and used the condition $\sum_{i}\dm_{ii}^{(0)} = 1$, which is required to determine $\dm^{(0)}$ uniquely. Therefore, upon normalizing Eq.~(\ref{eq:no30}) with help of Eq.~(\ref{eq:no31}) we obtain the first main result,
\begin{eqnarray}
\label{eq:no32}
\dm_{nn}^{(2)} & = & \sum_{i} S_{ni}S_{in} \left[V_{ni}\dm_{ii}^{(0)}-V_{in}\dm_{nn}^{(0)}+W_{in}^{\p\p}\frac{\partial\dm_{nn}^{(0)}}{\partial E_{n}}\right] \nonumber \\
& &-\dm_{nn}^{(0)}\sum_{i,j}S_{ji}S_{ij}W_{ij}^{\p\p}\frac{\partial\dm_{jj}^{(0)}}{\partial E_{j}}.
\end{eqnarray}
Using Eq.~(\ref{eq:no32}) to calculate the $2$-nd order diagonal elements we need to know the derivative of the $0$-th order reduced density matrix, $\partial\dm_{nn}^{(0)}/\partial E_{n}$. This derivative derives from Eq.~(\ref{eq:no21}), which is satisfied by $\dm^{(0)}$ and subsequently differentiate with respect to the energy $E_{n}$ to find
\begin{eqnarray}
\label{eq:no33}
\frac{\partial\dm_{nn}^{(0)}}{\partial E_{n}} & = & \frac{\sum_{\substack{i\ne n}}S_{ni}S_{in}\left(\frac{\partial \Wre_{ni}}{\partial \tf_{ni}}\dm_{ii}^{(0)}+\frac{\partial \Wre_{in}}{\partial \tf_{in}}\dm_{nn}^{(0)}\right)}{\sum_{\substack{i\ne n}}S_{ni}S_{in}\Wre_{in}}.
\end{eqnarray}
Therefore, we have all the ingredients at hand to calculate the $2$-nd order diagonal elements from Eq.~(\ref{eq:no32}): This constitutes the first main result of our work. In our derivation we have made no assumptions besides the validity of analytic continuation. The above outlined theory can be readily generalized to multiple heat baths; a topic to be addressed by us in future work\cite{Thingna1}.

The modified solution outlined above is correct not only up to $2$-nd order in system-bath coupling but additionally it is well suited for numerical studies: Numerical simulations with the RQME are very cumbersome because the relaxation tensor $\mathrm{R}^{(2)}$ in Eq.~(\ref{eq:no6}) scales as the fourth power\cite{Weiss} of the system Hilbert space dimension $N$. Therefore, in the steady state the computational complexity of the problem typically scales proportional to $N^{6}$, assuming that the analytic forms of the transition rates are known. On the other hand, in our modified Redfield solution all components of the reduced density matrix can be obtained by reference to the transition rates $W$ only, which scale as $N^{2}$. Thus, in the modified solution the computational complexity becomes drastically reduced to be of order $N^{3}$. This fact is equivalent to solving the quantum master equation with use of the continued fraction scheme\cite{Garcia}; it thus enables us to study systems with much larger Hilbert space dimension.
\section{Comparing modified Redfield solution with second order canonical perturbation theory }
\label{sec:4}
For a finite system-bath coupling the thermal equilibrium density matrix is typically no longer of Gibbs type (strict weak coupling limit) but rather of the generalized Gibbs form, $\dm^{\sm{\mathrm{eq}}} \propto \mathrm{Tr}_{\sm{\mathrm{B}}}(\e^{-\be H_{\sm{\mathrm{tot}}}})$, resulting in a quantum Hamiltonian of mean force\cite{Campisi}. It is interesting to know if this distribution can be obtained from a full non-Markovian dynamical theory of a system weakly coupled to a heat bath. Although this seems reasonable there is no agreed consensus on this issue from the viewpoint that the literature deals with a variety of perturbative quantum ($2$-nd order) master equations\cite{Romero,Geva2,Haake,Kubo1}.

Since the Redfield formalism is rigorously valid only in the $\ep \rightarrow 0$ limit it is expected that in this very limit the canonical form $\dm^{\sm{\mathrm{eq}}} \propto \e^{-\be H_{\sm{\mathrm{S}}}}$ emerges. In order to test the accuracy of our novel modified Redfield solution we implement an order by order comparison between canonical perturbation theory (CPT), which perturbatively expands the generalized Gibbs distribution, as detailed in the Appendix, with our modified Redfield solution. According to CPT (Eq.~(\ref{eq:noA.13}),~(\ref{eq:noA.14}), and~(\ref{eq:noA.15})) the reduced density matrix up to $2$-nd order in the system-bath coupling reads
\begin{eqnarray}
\dm_{nm}^{\sm{\mathrm{CPT}}} & = & \dm_{nm}^{(0),\sm{\mathrm{CPT}}} + \dm_{nm}^{(2),\sm{\mathrm{CPT}}}, \nonumber \\
\label{eq:no34}
\dm_{nm}^{(0),\sm{\mathrm{CPT}}} & = & \frac{\e^{-\be E_{n}}}{Z_{\sm{\mathrm{S}}}}\delta_{n,m}, \\
\label{eq:no35}
\dm_{nm}^{(2),\sm{\mathrm{CPT}}} & = &\frac{\sdm_{nm}}{Z_{\sm{\mathrm{S}}}}-\frac{\e^{-\be E_{n}}\sum_{i}\sdm_{ii}}{(Z_{\sm{\mathrm{S}}})^{2}}\delta_{n,m},
\end{eqnarray}
wherein the different contributions assume the form:
\begin{eqnarray}
Z_{\sm{\mathrm{S}}}&=&\sum_{l}\e^{-\be E_{l}}, \nonumber \\
\label{eq:no36}
\sdm_{nm} & = & \frac{1}{\tf_{mn}}\sum_{l}\left(\sdmt_{nl}S_{lm}-\sdmt_{ml}S_{ln}\right)~~~~(n \neq m), \nonumber \\
\sdmt_{nl} & = & S_{nl}\e^{-\be E_{n}}\left(\int_{0}^{\be}\drm x\ch(-\iu x)\e^{-x\tf_{ln}}-\frac{\dk_{\sm{0}}}{2}\right).\\
\label{eq:no37}
\sdm_{nn} & = & \sum_{l}\sdmtd_{nl}S_{ln}, \nonumber \\
\sdmtd_{nl} & = & S_{nl}\e^{-\be E_{n}}\Biggl[\be\left(\int_{0}^{\be}\drm x\ch(-\iu x)\e^{-x\tf_{ln}}-\frac{\dk_{\sm{0}}}{2}\right) \Biggr. \nonumber \\
&&\Biggl. -\int_{0}^{\be}\drm x \ch(-\iu x) x \e^{-x\tf_{ln}}\Biggr].
\end{eqnarray}

\subsection{Comparing the 0-th order result}
\label{subsec:4.1}
Let us first compare the $0$-th order reduced density matrix. For the harmonic baths described by Eq.~(\ref{eq:no3}) it can be shown that the bath-bath correlator $\ch(\tau)$ obeys the Kubo-Martin-Schwinger (KMS) condition\cite{Kubo1,Kubo2,Martin},
\begin{eqnarray}
\label{eq:no38}
\ch(-\tau) &=& \ch(\tau -\iu\be).
\end{eqnarray}
This implies that the real part of the transition rates $\Wre$ obey the detailed balance condition\cite{Lebowitz} given by,
\begin{eqnarray}
\label{eq:no39}
\Wre_{ij} & = &\e^{-\be\tf_{ij}} \Wre_{ji}.
\end{eqnarray}
The analytic form of the $0$-th order reduced density matrix can be obtained upon using Eq.~(\ref{eq:no21}) as,
\begin{equation}
\label{eq:no40}
\dm_{nm}^{(0)} = \frac{\e^{-\be E_{n}}}{Z_{\sm{S}}}\delta_{n,m},
\end{equation}
where $Z_{\sm{\mathrm{S}}}=\sum_{l}\e^{-\be E_{l}}$. A direct comparison between Eq.~(\ref{eq:no40}) and Eq.~(\ref{eq:no34}) yields the expected result that at the $0$-th order CPT agrees with our modified, $0$-th order Redfield solution.
\subsection{Comparing the 2-nd order result}
\label{subsec:4.2}
More intriguing is the comparison of the modified Redfield solution with the $2$-nd order CPT-result. The $2$-nd order reduced density matrix obtained from CPT can be manipulated further so that it indeed matches precisely our modified Redfield solution. In order to demonstrate this fact we first simplify the integral occurring in $\sdmt$, Eq.~(\ref{eq:no36}), by using the definition of the bath-bath correlator $\ch(\tau)$ in Eq.~(\ref{eq:no11}) to obtain
\begin{eqnarray}
\label{eq:no41}
\int_{0}^{\be}&&\drm x\ch(-\iu x)\e^{-x\tf_{ij}} = \nonumber \\
& & -\int_{0}^{\infty}\frac{\drm\w}{\pi} \sd(\w) \left( \frac{n_{\w}}{\w-\tf_{ij}} - \frac{\left(n_{\w}+1\right)}{\w+\tf_{ij}}\right) \nonumber \\
&&-\frac{\e^{-\be\tf_{ij}}}{\pi}\int_{0}^{\infty}\drm\w \sd(\w) \left( \frac{n_{\w}}{\w+\tf_{ij}} - \frac{\left(n_{\w}+1\right)}{\w-\tf_{ij}}\right), \nonumber \\
\end{eqnarray}
where we have interchanged the $\w$- (stemming from $\ch(\tau)$) and $x$- integration and performed the $x$-integral analytically. We next express the right hand side in terms of the transition rates $\tilde{W}$, which enter in our modified Redfield solution. In order to do this we use the so termed Sokhotskyi-Plemelj formula\cite{Kress},
\begin{eqnarray}
\label{eq:no42}
\int_{0}^{\infty} \e^{\pm \iu \Omega \tau} \drm\tau &=& \pi \delta(\Omega) \pm \iu \mathrm{P}\left(\frac{1}{\Omega}\right).
\end{eqnarray}
Here, P denotes the principal value. Therefore, using the above identity along with Eq.~(\ref{eq:no8}) and Eq.~(\ref{eq:no11}) we can express the imaginary part of the transition rates $\Wim$ in the form
\begin{eqnarray}
\label{eq:no43}
\Wim_{ij}&=&\mathrm{P}\int_{0}^{\infty}\frac{\drm\w}{\pi} \sd(\w) \left( \frac{n_{\w}}{\w-\tf_{ij}} - \frac{\left(n_{\w}+1\right)}{\w+\tf_{ij}}\right).
\end{eqnarray}
Therefore using the above equation, Eq.~(\ref{eq:no41}) can be expressed as,
\begin{eqnarray}
\label{eq:no44}
-\int_{0}^{\be}\drm x \e^{-x\tf_{ij}}\ch(-\iu x) & = &\Wim_{ij}+\e^{-\be\tf_{ij}}\Wim_{ji}.
\end{eqnarray}
\subsubsection{Off-diagonal elements}
\label{subsubsec:4.2.1}
Upon use of Eq.~(\ref{eq:no44}) the $2$-nd order off-diagonal elements from CPT, i.e., Eq.~(\ref{eq:no35}) can be expressed in terms of $\Wim$ as
\begin{eqnarray}
\label{eq:no45}
\dm_{nm}^{(2),\sm{\mathrm{CPT}}} & = & \frac{1}{\tf_{nm}}\sum_{i}S_{ni}S_{im}\left[\frac{\e^{-\be E_{i}}}{Z_{\sm{S}}}\left(W_{ni}^{\p\p}-W_{mi}^{\p\p}\right)\right.\nonumber \\
&&+\left.\frac{\e^{-\be E_{n}}}{Z_{\sm{S}}}W_{in}^{\p\p}-\frac{\e^{-\be E_{m}}}{Z_{\sm{S}}}W_{im}^{\p\p}\right],
\end{eqnarray}
where we have absorbed the $\dk_{\sm{0}}$ into $W^{\p\p}$, according to Eq.~(\ref{eq:no7}). Formally adding the real part of the transition rates $W^{\p}$ into Eq.~(\ref{eq:no45}), but noting that this so added contributions vanish identically by virtue of detailed balance in Eq. (\ref{eq:no39}), we find the result
\begin{eqnarray}
\label{eq:no46}
\dm_{nm}^{(2),\sm{\mathrm{CPT}}} & = & \frac{1}{\iu\tf_{nm}} \sum_{i} S_{ni}S_{im} \left[ \left(W_{ni}+W_{mi}^{*}\right)\frac{\e^{-\be E_{i}}}{Z_{\sm{S}}}\right. \nonumber\\
& & \left.-W_{in}^{*}\frac{\e^{-\be E_{n}}}{Z_{\sm{S}}}-W_{im}\frac{\e^{-\be E_{m}}}{Z_{\sm{S}}}\right], (n\neq m).
\end{eqnarray}
Upon comparing Eq.~(\ref{eq:no22}) with Eq.~(\ref{eq:no46}) we find that the CPT and our modified Redfield solution are identical.
\subsubsection{Diagonal elements}
\label{subsubsec:4.2.2}
Most importantly, we next test the agreement between the $2$-nd order diagonal elements from CPT with our modified Redfield solution. Noting that the integral occurring in Eq.~(\ref{eq:no35}) is the derivative of Eq.~(\ref{eq:no44}) w.r.t $\tf_{ij}$ we obtain,
\begin{eqnarray}
\label{eq:no47}
\dm_{nn}^{(2),\sm{\mathrm{CPT}}} & = & \sum_{i}S_{ni}S_{in}\left(\frac{\e^{-\be E_{i}}}{Z_{\sm{S}}}V_{ni}-\frac{\e^{-\be E_{n}}}{Z_{\sm{S}}}V_{in}\right) \nonumber \\
& &-\be\frac{\e^{-\be E_{n}}}{Z_{\sm{S}}}\Biggl[\sum_{i}S_{ni}S_{in}\Wim_{in} \Biggr. \nonumber \\
& &\Biggl.-\sum_{i,l}S_{li}S_{il}\frac{\e^{-\be E_{l}}}{Z_{\sm{S}}}\Wim_{il}\Biggr],
\end{eqnarray}
where $V_{ij}$ has been defined in Eq.~(\ref{eq:no27}). Because $\partial\dm_{ii}^{(0)}/\partial E_{i} = -\be\dm_{ii}^{(0)}$, Eq.~(\ref{eq:no30}) exactly matches Eq.~(\ref{eq:no47}). This constitutes a second main result: Namely CPT up to $2$-nd order and our modified Redfield solution are indeed in perfect agreement. This shows that in the weak, but finite coupling limit the long-time thermal reduced density matrix stemming from a non-Markovian theory is of the generalized Gibbs form.
\section{A test case: Damped Harmonic Quantum Oscillator}
\label{sec:5}
In this section we compare our modified Redfield solution to the \emph{exact} nonequilibrium Green's function (NEGF) results\cite{Dhar} for a damped harmonic oscillator that is linearly coupled to a thermal heat bath. This comparison will allow us to estimate the system-bath coupling strengths that can be probed safely by employing the presented modified scenario. In order to do so we use a specific spectral density $\sd(\w)$ of the heat bath. Several phenomenological forms of the spectral density are in use in the literature for the damped oscillator quantum dynamics; sometimes a numerical decomposition is employed to save computational costs\cite{Tanimura2,Meier}. Here we use the common Lorentz-Drude (LD) spectral density,

\begin{eqnarray}
\label{eq:no48}
\sd(\w) & = & \frac{M\dk \w}{1+\left(\w/\w_{\sm{\mathrm{D}}}\right)^{2}},
\end{eqnarray}
where $\w_{\sm{\mathrm{D}}}$ denotes the cutoff frequency and $\dk$ is the phenomenological Stokesian damping coefficient which characterizes the system-bath coupling strength. Since $\dk \propto \sum_{n=1}^{\infty} c_{n}^{2}$, the spectral density is of $2$-nd order in system-bath coupling. Decomposing the hyperbolic cotangent in Eq.~(\ref{eq:no11}) into its Matsubara frequencies, $\nu_{l}=2\pi l T$, where $T$ denotes the temperature of the bath, and noting that the resultant equation exhibits poles at $\w = \pm \iu \w_{\sm{\mathrm{D}}}$ and $\w = \pm \iu \nu_{l}$ we can calculate $\ch(\tau)$ explicitly by use of the residue theorem to obtain
\begin{eqnarray}
\label{eq:no49}
\ch(\tau)&=&\frac{M\dk}{2}\w_{\sm{\mathrm{D}}}^{2}\mathrm{cot}\left(\frac{\be \w_{\sm{\mathrm{D}}}}{2}\right)\e^{-\w_{\sm{\mathrm{D}}}\tau}-\frac{2M\dk}{\be}\sum_{l=1}^{\infty}
\frac{\nu_{l}\e^{-\nu_{l}\tau}}{1-(\nu_{l}/\w_{\sm{\mathrm{D}}})^{2}} \nonumber \\
&&-\iu\frac{M\dk}{2}\w_{\sm{\mathrm{D}}}^{2}\e^{-\w_{\sm{\mathrm{D}}}\tau}\rm{sgn}(\tau).
\end{eqnarray}
Therefore, the components of the transition rates $W$, defined in Eq.~(\ref{eq:no8}), read
\begin{eqnarray}
\label{eq:no50}
\Wre_{ij} &=&\frac{M\dk \w_{\sm{\mathrm{D}}}^{2}}{2(\w_{\sm{\mathrm{D}}}^{2}+\tf_{ij}^{2})}\left[\w_{\sm{\mathrm{D}}}\mathrm{cot}\left(\frac{\be \w_{\sm{\mathrm{D}}}}{2}\right)-\tf_{ij}\right] \nonumber \\
&&-\frac{2M\dk}{\be}\sum_{l=1}^{\infty}\frac{\nu_{l}^{2}}{(1-(\nu_{l}/\w_{\sm{\mathrm{D}}})^{2})(\nu_{l}^{2}+\tf_{ij}^{2})},\\
\label{eq:no51}
\Wim_{ij}&=&\frac{M\dk \w_{\sm{\mathrm{D}}}^{2}\tf_{ji}}{2(\w_{\sm{\mathrm{D}}}^{2}+\tf_{ij}^{2})}\left[\mathrm{cot}\left(\frac{\be \w_{\sm{\mathrm{D}}}}{2}\right)+\frac{\w_{\sm{\mathrm{D}}}}{\tf_{ij}}\right] \nonumber \\
&& +\frac{2 M\dk \tf_{ij}}{\be}\sum_{l=1}^{\infty}\frac{\nu_{l}}{[1-(\nu_{l}/\w_{\sm{\mathrm{D}}})^{2}](\nu_{l}^{2}+\tf_{ij}^{2})},
\end{eqnarray}
with the relation that
\begin{equation}
\label{eq:no52}
\dk_{\sm{0}} = \dk\w_{\sm{\mathrm{D}}}.
\end{equation}

The system Hamiltonian is a single harmonic oscillator, reading
\begin{eqnarray}
\label{eq:no53}
H_{\sm{\mathrm{S}}} = \frac{p^{2}}{2M}+\frac{1}{2}M\w_{\sm{0}}^{2}x^2,
\end{eqnarray}
where $x$, $p$, $M$, and $\w_{\sm{0}}$ are the position, momentum, mass and angular frequency of the oscillator. The harmonic oscillator is linearly coupled to the bath via the $x$-coordinate. This implies that $S=x$ in Eq.~(\ref{eq:no5}). Throughout this work the system-bath coupling will be measured in a dimensionless parameter, defined by taking the ratio between the damping coefficient $\dk$ and the angular oscillator frequency, i.e., $\dk/\w_{\sm{0}}$. Since the Redfield formalism is formulated in terms of eigenbasis of the system Hamiltonian of finite dimension, we choose a system Hilbert space that is sufficiently large so that even at the highest temperatures the occupation probability of finding the particle in the highest available energy levels is practically zero. We do this by iteratively increasing the size of the system Hilbert space until at least five largest energy levels possess a population less than $10^{-15}$: In our case of the damped harmonic oscillator this results in around $40$ levels. Using these $40$ levels we can cover a temperature range up to five times the Debye temperature, $T_{\sm{\mathrm{D}}} = (\h\w_{\sm{0}})/k_{\sm{\mathrm{B}}}$.
\begin{figure}
\begin{center}
\includegraphics[scale=0.35]{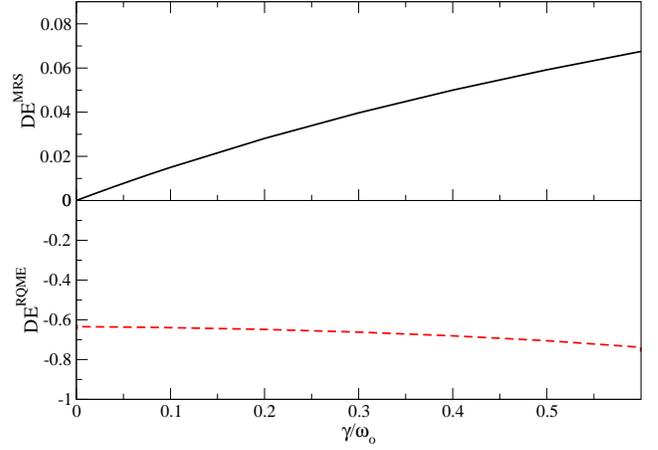}
\end{center}
\caption{(Color Online) Plot of the discrepancy error DE$^{\mathrm{X}}$ of the ground state population versus the dimensionless system-bath coupling strength ($\dk/\w_{\sm{0}}$) for a damped harmonic quantum oscillator. Top panel: The (black) solid line depicts the rather small discrepancy for our \emph{modified} Redfield solution ($X=$ MRS) and the bottom panel (red) dashed line shows the large discrepancy obtained via the ordinary Redfield quantum master equation ($X=$ RQME). Our parameters used for the calculation are $M = 1~\mathrm{u}$, $\w_{\sm{0}} = 1.3\times10^{14}~\mathrm{Hz}$, $T = 50~\mathrm{K}$ and the cutoff is chosen at $\w_{\sm{\mathrm{D}}} = 10 ~\w_{\sm{0}}$.}
\label{fig:1}
\end{figure}

The main goal in this work is to correctly evaluate the $2$-nd order diagonal elements. At the $0$-th order level the RQME and our modified Redfield solution give the canonical solution, which matches the result obtained from the NEGF method by taking the zero coupling limit. Therefore, in order to sensitively compare the $2$-nd order elements we define a relative discrepancy error $\mathrm{DE}^{\mathrm{X}}$ as follows:
\begin{eqnarray}
\label{eq:no54}
\mathrm{DE}^{\mathrm{X}} &= &\frac{\dm^{\sm{\mathrm{NEGF}}}-\dm^{\sm{\mathrm{X}}}}{(\dk/\w_{\sm{0}})}\;,
\end{eqnarray}
where $\dm^{\sm{\mathrm{NEGF}}}$ denotes the exact reduced density matrix obtained from NEGF method, $\dm^{\sm{\mathrm{X}}}$ is the reduced density matrix obtained from the perturbative method; being either our modified Redfield solution ($\mathrm{X} =$ MRS) or the Redfield quantum master equation ($\mathrm{X} =$ RQME), and the ratio $\dk/\w_{\sm{0}}$ specifies the overall system-bath coupling strength. Since the $4$-th order term of the reduced density matrix is of the order of $(\dk/\w_{\sm{0}})^{2}$, it is expected that the discrepancy error is one order lower, i.e., $O(\dk/\w_{\sm{0}})$ if and only if the $2$-nd order elements are calculated correctly. In order to check this behavior we plot the discrepancy error versus $\dk/\w_{\sm{0}}$ for the ground state population at $T = 50~\mathrm{K}$ in Fig.~\ref{fig:1}.

Since the temperature is chosen low the population $\dm_{11}$ presents an appropriate quantifier for the complete reduced density matrix. The figure depicts that the discrepancy error for our modified Redfield solution (solid black line) indeed stays throughout of the order of ($\dk/\w_{\sm{0}}$) for all coupling strengths $\dk/\w_{\sm{0}}$; in contrast, for the RQME (dashed red line) the discrepancy error grows in absolute value $\gg O(\dk/\w_{\sm{0}})$, indicating the inaccuracy in the $2$-nd order elements. In the limit ($\dk/\w_{\sm{0}} \rightarrow 0$) the discrepancy error should vanish: This holds true only for our modified Redfield solution whereas the Redfield solution depicts a finite value which indicates that this $2$-nd order (i.e., being proportional $c_n^2$) solution indeed is not correct to leading $2$-nd order. The temperature does not play a major role, since all the features in the discrepancy error remain the same up to temperatures of $\sim3000$ K. Therefore, for all values of the weak-to-moderate system-bath coupling strengths our modified Redfield solution is able to predict the $2$-nd order elements correctly, whereas the RQME fails to do so already for small values of ($\dk/\w_{\sm{0}}$).

\begin{figure}
\begin{center}
\includegraphics[scale=0.33]{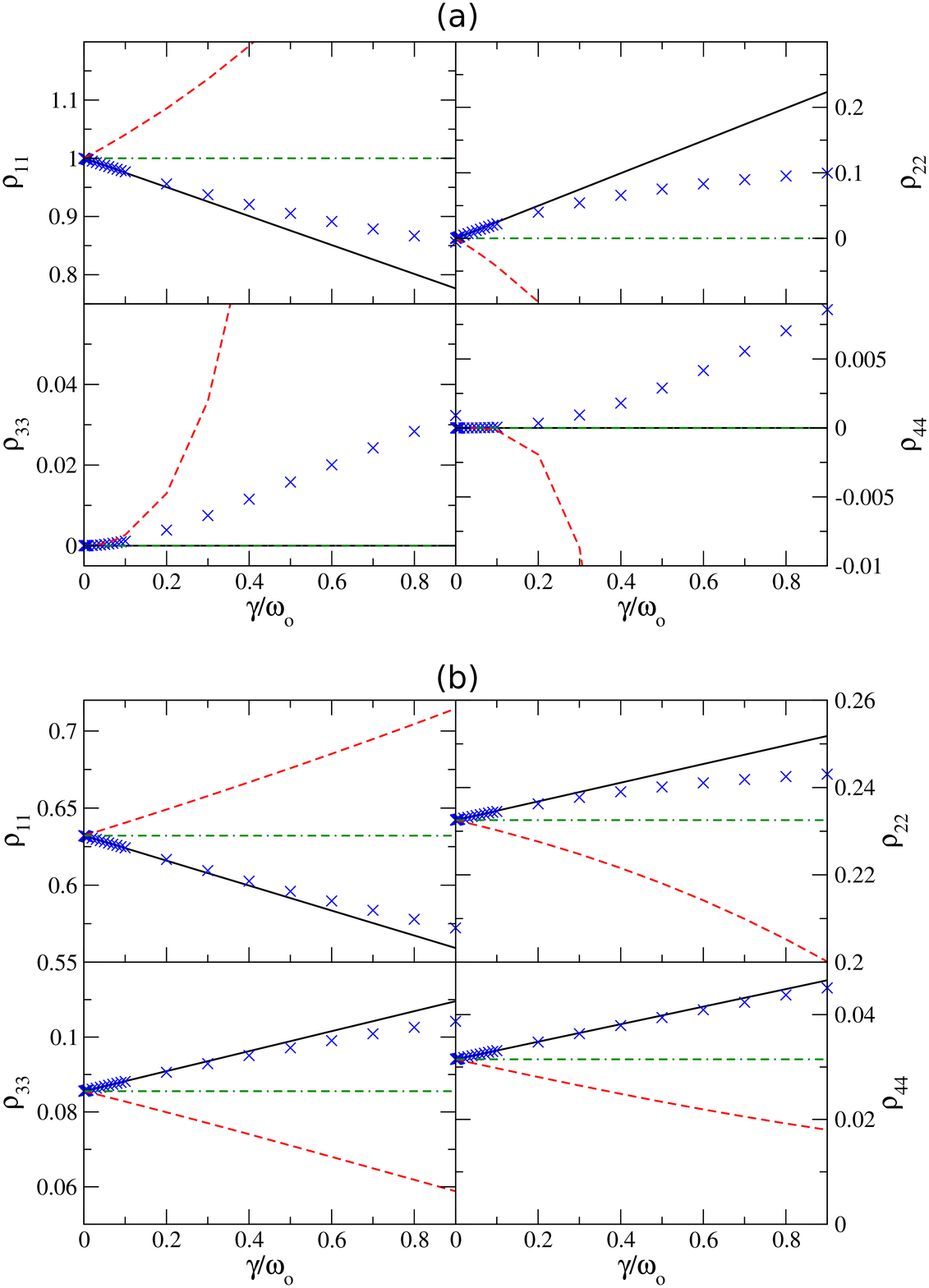}
\end{center}
\caption{(Color Online) Graph of the populations for the first four lowest lying energy levels versus the dimensionless system-bath coupling strength $\dk/\w_{\sm{0}}$ for a damped quantum harmonic oscillator. The (black) solid lines correspond to our modified Redfield solution, the (red) dashed lines present the results for the Redfield quantum master equation, the (green) dotted-dashed lines depict the results for the Lindblad solution, while the (blue) crosses represent the exact result using NEGF. Panel (a) is for the temperature of $T = 50~\mathrm{K}$ and panel (b) corresponds to a temperature of $T = 1000~\mathrm{K}$. The remaining parameters used for the calculation are $M = 1~\mathrm{u}$ , $\w_{\sm{0}} = 1.3\times10^{14}~\mathrm{Hz}$ and $\w_{\sm{\mathrm{D}}} = 10 ~\w_{\sm{0}}$.}
\label{fig:2}
\end{figure}

In Fig.~\ref{fig:2} we study the actual population values for the first few levels as a function of $\dk/\w_{\sm{0}}$ for two different temperature. Fig.~\ref{fig:2}(a) corresponds to a temperature of $T = 50~\mathrm{K}$ and \ref{fig:2}(b) is for $T = 1000~\mathrm{K}$. We have opted to plot these two extreme temperatures because for all intermediate temperatures values the features of the plot remain practically the same. We compare our modified Redfield solution (black solid line) with the RQME (dashed red line), NEGF results (blue crosses) and the Lindblad master equation (dotted green line). The Lindblad master equation is extensively used in the literature\cite{Breuer, Weiss} due to its ease in computation and its preservation of positivity. Although positivity is an essential criteria, the Lindblad solution for the damped quantum harmonic oscillator case is the canonical distribution\cite{Honda} with no explicit dependence on coupling strength. Put differently, the Lindblad solution always fails to capture the effects of finite system-bath coupling. On the other hand the RQME depicts severe deviations from the exact result for small, but finite coupling strengths.

In the extreme low temperature regime the RQME yields negative populations, note the results for ($\dm_{22}, \dm_{44}$) in panel Fig.~\ref{fig:2}(a) already for weak coupling strengths, indicating that the validity of the solution holds only in the zero coupling limit. The steady state solution of the RQME has been critiqued before\cite{Munro,Blanga} for producing unphysical, negative populations. We can now assess that the reason for its breakdown is rooted in the incorrect $2$-nd order diagonal elements. The modified Redfield solution matches the exact solution quite well for system-bath couplings $\dk/\w_{\sm{0}}$ as strong as 0.2, even at low temperatures. At high temperatures our modified solution yields a most impressive agrement with the exact results, extending over sizable regimes of coupling strengths up to $\dk/\w_{\sm{0}} \gtrsim 0.6$. Beyond a coupling strength $\dk/\w_{\sm{0}} \sim 0.6$ the modified Redfield density matrix is no longer positive definite, which is determined upon examining the eigenvalues of the reduced density matrix; this indicates a breakdown of $2$-nd order perturbation theory beyond this value. Nevertheless, the presented modified Redfield solution provides a decisive and salient improvement over the RQME in that the coupling strengths that can be probed accurately becomes sizable.
\section{Concluding remarks and outlook}
\label{sec:6}
In this paper we have demonstrated via the exact comparison with canonical perturbation theory and extensive numerics that the Redfield quantum master equation is inaccurate for the steady state. This failure is the result of incorrect second order diagonal elements. This is in the spirit of remarks made before by Fleming \emph{et~al.}\cite{Fleming2}. Their suggestion in overcoming this flaw by the use of the fourth order tensor to improve the second order accuracy is numerically extremely cumbersome. Their suggestion to use instead the Davies approximation\cite{Davies1,Davies2,Davies3,Dumcke} in order to obtain the thermal steady state reduced density matrix correct up to second order is not appropriate either. This is so because in case of the Davies approximation the long-time limit can be taken only if we take $\ep \rightarrow 0$, so that the product $\ep^{2} t$ remains constant\cite{Dumcke}. This in turn immediately implies that the Davies approximation is precise only to $0$-th order accuracy in the long-time limit, where it agrees with the Lindblad solution. Attempts have been made to correctly evaluate the second order diagonal elements using the Dyson expansion\cite{Thingna2}: In this context it must be noted, however, because the Dyson series is \emph{asymptotically divergent}\cite{Dyson}, and although the off-diagonals in fact agree, the second order diagonal elements are found \emph{not} to match the exact result of Dhar \emph{et~al.}\cite{Dhar} for the damped harmonic oscillator problem.

Therefore, since most of the perturbative methods fail to capture the effects of finite system-bath coupling in the long-time limit, we put forward a modified solution to the Redfield quantum master equation which reproduces the second order elements exactly. The derivation is based on obtaining the second order diagonal elements from the off-diagonal ones using an analytic continuation procedure as detailed in Sec.~\ref{subsec:3.2}. The result of this scheme is unique indicating a unique steady state for the reduced density matrix. In order to test the validity of our solution, we have compared our modified Redfield solution to canonical perturbation theory and demonstrated that our modified solution agrees with the generalized Gibbs distribution up to second order in the system-bath coupling strength for a general system that is coupled to a harmonic oscillator bath. This indicates that even in the weak, but finite, system-bath coupling limit the system thermalizes to a generalized Gibbs distribution. As will be elaborated elsewhere our method is also applicable to systems connected with multiple-baths, thus exhibiting nonequilibrium steady state transport\cite{Thingna1}.

As an illustrative example we tested and compared in Sec.~\ref{sec:5} the reduced density matrix obtained by our modified solution, the Redfield formalism, and the Lindblad master equation against the exact NEGF results for a damped harmonic oscillator. We find that our modified solution agrees quite well with the exact result for coupling strengths as strong as $\dk/\w_{\sm{0}} = 0.2$, showing a major improvement over the RQME which matches the exact result only in the limit $\dk/\w_{\sm{0}} \rightarrow 0$. On the other hand the Lindblad solution for the damped oscillator case always yields the canonical distribution, wrongly indicating that the solution is not affected by the system-bath coupling strength.

The presented modified Redfield solution further is numerically very efficient; this is mainly so because with our scenario the computational complexity scales as $N^{3}$, where $N$ is the system Hilbert space dimension, as compared to $N^{6}$ for the Redfield formalism. This fact allows us to describe accurately not only the effects of finite system-bath coupling, but as well as to explore systems with rather large Hilbert space dimensions. A yet unsolved challenge consists in the extension of our scheme to the \emph{time-dependent} relaxation of the reduced density matrix $\dm(t)$ and, in this context, also the extension to study the differing relaxation processes that stem from different initial preparation schemes away from the typically used case of a correlation-free initial preparation. Assuming bath spectral densities that assure an ergodic behavior, the long-time limit is not affected by the initial preparation, being in distinct contrast to its temporal relaxation. Yet another unsolved objective presents the perturbative, accurate study of multi-time correlations of open system observables, both time-homogeneous thermal and time-dependent nonequilibrium correlations beyond the weak coupling limit.
\section*{Acknowledgements}
\label{sec:7}
We would like to thank J. L. Garc\'{\i}a-Palacios and Meng Lee Leek for discussions on the accuracy of the Redfield master equation, Lee Chee Kong for helpful discussions on canonical perturbation theory, Adam Zaman Chaudhry on the implementation of the master equation, and Bijay Kumar Agarwalla and Lifa Zhang for insightful discussions. This work has been partly supported (J.-S. W.) by an URC grant R-144-000-257-112 and (P.H.) via the support with the German Excellence Initiative ``Nanosystems Initiative Munich (NIM)''.
\renewcommand{\theequation}{A\arabic{equation}}
\setcounter{equation}{0}
\section*{Appendix: Canonical Perturbation Theory}
\label{append:A}
With this Appendix we outline the basic reasoning underlying canonical perturbation theory\cite{Romero,Geva2,Mori} (CPT). This will assist us in determining the correct equilibrium reduced density matrix up to second order in coupling strength for a harmonic bath which is coupled bi-linearly to a general system. The basic idea dates back to the works of Peierls\cite{Peierls} and Landau\cite{Landau} who calculated the free energy of the full system using a similar expansion. Here we employ similar techniques for the reduced density matrix, which in case of the equilibrium problem is well defined by the generalized Gibbs distribution\cite{Campisi}:
\begin{equation}
\label{eq:noA.1}
\dm^{\mathrm{eq}} = \frac{\mathrm{Tr}_{\mathrm{B}} \e^{-\be H_{\sm{\mathrm{tot}}}}}{\mathrm{Tr}\e^{-\be H_{\sm{\mathrm{tot}}}}},
\end{equation}
where $H_{\sm{\mathrm{tot}}}$ is defined in Eq.~(\ref{eq:no1}). We now use the Kubo identity\cite{Kubo1},
\begin{equation}
\e^{\be(A+B)}=\e^{\be A}\left[ \id + \int_{0}^{\be} \drm\lambda \e^{-\lambda A}B\e^{\lambda(A+B)}\right],
\label{eq:noA.2}
\end{equation}
which is exact. Upon expanding $\e^{-\be H_{\sm{\mathrm{tot}}}}$ up to second order in the coupling strength and taking the trace over the bath degrees of freedom we obtain
\begin{eqnarray}
\label{eq:noA.3}
\mathrm{Tr}_{\mathrm{B}}&& (\e^{-\be H_{\sm{\mathrm{tot}}}})  =  \nonumber \\
&&\e^{-\be H_{\sm{\mathrm{S}}}}\Biggl[\id-\frac{\dk_{\sm{0}}}{2}\int_{0}^{\be}\drm\be_{1}\tilde{S}(-\iu\be_{1})\tilde{S}(-\iu\be_{1})\Biggr. \nonumber \\
&&\Biggl.+ \int_{0}^{\be}\drm\be_{1}\int_{0}^{\be_{1}}\drm\be_{2}\tilde{S}(-\iu\be_{1})\tilde{S}(-\iu\be_{2})\ch(-\iu(\be_{1}-\be_{2}))\Biggr], \nonumber \\
\end{eqnarray}
where $\tilde{S}(-\iu\be_{1})=\e^{\be_{1}H_{\sm{\mathrm{S}}}}S\e^{-\be_{1}H_{\sm{\mathrm{S}}}}$ is the free evolving system operator in imaginary time and $\ch(-\iu(\be_{1}-\be_{2}))$ is the imaginary-time bath correlator as defined in Sec.~\ref{sec:2}. Using Eq.~(\ref{eq:noA.3}) in Eq.~(\ref{eq:noA.1}) the CPT reduced density matrix thus reads
\begin{eqnarray}
\label{eq:noA.4}
\dm^{\sm{\mathrm{CPT}}} & = & \frac{\e^{-\be H_{\sm{\mathrm{S}}}}}{Z_{\sm{\mathrm{S}}}}+\frac{\sdm}{Z_{\sm{\mathrm{S}}}}-\frac{\e^{-\be H_{\sm{\mathrm{S}}}}\mathrm{Tr}_{\sm{\mathrm{S}}}(\sdm)}{(Z_{\sm{\mathrm{S}}})^{2}},
\end{eqnarray}
where,
\begin{eqnarray}
\label{eq:noA.5}
\sdm & = & \int_{0}^{\be}\drm\be_{1}\int_{0}^{\be_{1}}\drm\be_{2}\tilde{S}(-\iu\be_{1})\tilde{S}(-\iu\be_{2})\ch(-\iu(\be_{1}-\be_{2}))\nonumber \\
&&-\frac{\dk_{\sm{0}}}{2}\int_{0}^{\be}\drm\be_{1}\tilde{S}(-\iu\be_{1})\tilde{S}(-\iu\be_{1}),\\
\label{eq:noA.6}
Z_{\sm{\mathrm{S}}} & = & \mathrm{Tr}_{\sm{\mathrm{S}}}(\e^{-\be H_{\sm{\mathrm{S}}}}).
\end{eqnarray}
Next writing Eq.~(\ref{eq:noA.4}) in the basis of the system Hamiltonian we obtain,
\begin{eqnarray}
\label{eq:noA.7}
\dm_{nm}^{\sm{\mathrm{CPT}}} & = & \frac{\e^{-\be E_{n}}}{Z_{\sm{\mathrm{S}}}}\delta_{n,m}+\frac{\sdm_{nm}}{Z_{\sm{\mathrm{S}}}}-\frac{\e^{-\be E_{n}}\sum_{i}\sdm_{ii}}{(Z_{\sm{\mathrm{S}}})^{2}}\delta_{n,m}, \nonumber \\
\end{eqnarray}
wherein
\begin{eqnarray}
\label{eq:noA.8}
\sdm_{nm}&=&\sum_{l}S_{nl}S_{lm}\e^{-\be \en_{n}} \nonumber \\
&&\left[\int_{0}^{\be}\drm\be_{1}\e^{\be_{1}\tf_{nl}}\int_{0}^{\be_{1}}\drm\be_{2}\e^{\be_{2}\tf_{lm}}\ch(-\iu(\be_{1}-\be_{2}))\right. \nonumber \\
& &\left.-\frac{\dk_{\sm{0}}}{2}\int_{0}^{\be}\drm\be_{1}\e^{\be_{1}\tf_{nm}}\right].
\end{eqnarray}
In Eq.~(\ref{eq:noA.8}) $\tf_{nm}=E_{n}-E_{m}$ has the same definition as in Eq.~(\ref{eq:no8}).

The main task in CPT is to evaluate the elements of the matrix $\sdm$, Eq.~(\ref{eq:noA.8}). In order to do this we
split the matrix $\sdm$ into its diagonal and off-diagonal elements and deal with each part separately, as detailed below.

\subsection {Off-diagonal elements of the matrix $\sdm_{nm}$}

In order to obtain the off-diagonal elements of the matrix $\sdm$ we make the following change of variables: $x = \be_{1}-\be_{2},y = \be_{1}+\be_{2}$ and then we perform the $y$ integral analytically to find
\begin{eqnarray}
\label{eq:noA.9}
\sdm_{nm} & = & \frac{1}{\tf_{mn}}\sum_{l}\left(\sdmt_{nl}S_{lm}-\sdmt_{ml}S_{ln}\right),
\end{eqnarray}\\
where,
\begin{eqnarray}
\label{eq:noA.10}
\sdmt_{nl} & = & S_{nl}\e^{-\be E_{n}}\left(\int_{0}^{\be}\drm x\ch(-\iu x)\e^{-x\tf_{ln}}-\frac{\dk_{\sm{0}}}{2}\right). \nonumber \\
\end{eqnarray}

\subsection {Diagonal elements of the matrix $\sdm_{nn}$}
For the diagonal elements of $\sdm$, by using the same set of transformations as before, the integrals simplify and
the diagonal elements of matrix $\sdm$ emerge as
\begin{eqnarray}
\label{eq:noA.11}
\sdm_{nn} & = & \sum_{l}\sdmtd_{nl}S_{ln},
\end{eqnarray}
where,
\begin{eqnarray}
\label{eq:noA.12}
\sdmtd_{nl} & = & S_{nl}\e^{-\be E_{n}}\Biggl[\be\left(\int_{0}^{\be}\drm x\ch(-\iu x)\e^{-x\tf_{ln}}-\frac{\dk_{\sm{0}}}{2}\right) \Biggr. \nonumber \\
&&\Biggl. -\int_{0}^{\be}\drm x \ch(-\iu x) x \e^{-x\tf_{ln}}\Biggr].
\end{eqnarray}

In summary, the thermal equilibrium reduced density matrix obtain via CPT is given, up to second order, by the generalized Gibbs state, reading:
\begin{eqnarray}
\label{eq:noA.13}
\dm_{nm}^{\sm{\mathrm{CPT}}} & = & \dm_{nm}^{(0),\sm{\mathrm{CPT}}} + \dm_{nm}^{(2),\sm{\mathrm{CPT}}}
\end{eqnarray}
where,
\begin{eqnarray}
\label{eq:noA.14}
\dm_{nm}^{(0),\sm{\mathrm{CPT}}} & = & \frac{\e^{-\be E_{n}}}{Z_{\sm{\mathrm{S}}}}\delta_{n,m}, \\
\label{eq:noA.15}
\dm_{nm}^{(2),\sm{\mathrm{CPT}}} & = &\frac{\sdm_{nm}}{Z_{\sm{\mathrm{S}}}}-\frac{\e^{-\be E_{n}}\sum_{i}\sdm_{ii}}{(Z_{\sm{\mathrm{S}}})^{2}}\delta_{n,m}.
\end{eqnarray}
Here, the off-diagonal elements of $\sdm_{nm}$ are given by Eq.~(\ref{eq:noA.9}) and the diagonal elements are
given by Eq.~(\ref{eq:noA.11}),~(\ref{eq:noA.12}). Eq.~(\ref{eq:noA.13}) exhibits that the equilibrium reduced density matrix obtained via CPT is hermitian and is normalized properly with trace over the system degrees of freedom equal to $1$.


\begin{thebibliography}{99}
\bibitem{Nakajima} S. Nakajima, Prog. Theor. Phys. \textbf{20}, 948 (1958).
\bibitem{Zwanzig1} R. Zwanzig, J. Chem. Phys. \textbf{33}, 1338 (1960).
\bibitem{Fulinski} A. Fulinski, and W. J. Kramarczyk, Physica \textbf{39}, 575 (1968).
\bibitem{Tokuyama} M. Tokuyama, H. Mori, Prog. Theor. Phys. \textbf{55}, 411 (1976).
\bibitem{Hanggi1} P. H\"anggi, and H. Thomas, Z. Phys. B -- Con. Mat. \textbf{26}, 85 (1977).
\bibitem{Grabert1} H. Grabert, P. Talkner, and P. H\"anggi, Z. Phys. B -- Con. Mat. \textbf{26}, 389 (1977).
\bibitem{Shibata} F. Shibata, Y. Takahashi, and N. Hashitsume, J. Stat. Phys. \textbf{17}, 171 (1977).
\bibitem{Nan} G. Nan, Q. Shi, and Z. Shuai, J. Chem. Phys. \textbf{130}, 134106 (2009).
\bibitem{Tanimura} Y. Tanimura, J. Phys. Soc. Jpn. \textbf{75}, 082001 (2006); and references therein.
\bibitem{Yan} Y. Yan, F. Yan, Y. Liu and J. Shao, Chem. Phys. Lett. \textbf{395}, 216 (2004).
\bibitem{Shao} J. Shao, J. Chem. Phys. \textbf{120}, 5053 (2004).
\bibitem{Xu} R.-X. Xu, P. Cui, X.-Q. Li, Y. Mo, and Y. J. Yan, J. Chem. Phys. \textbf{122}, 041103 (2005).
\bibitem{Grabert2} H. Grabert, U. Weiss, and P. H\"anggi, Phys. Rev. Lett. \textbf{52}, 2193 (1984).
\bibitem{Grabert3} H. Grabert, U. Weiss, and P. Talkner, Z. Phys. B -- Con. Mat. \textbf{55}, 87 (1984).
\bibitem{Grabert4} H. Grabert, P. Schramm, and G. L. Ingold, Phys. Rep. \textbf{168}, 115 (1988).
\bibitem{Riseborough} P. Riseborough, P. H\"anggi, and U. Weiss, Phys. Rev. A \textbf{31}, 471 (1985).
\bibitem{Hu} B. L. Hu, J. P. Paz, and Y. Zhang Phys. Rev. D \textbf{45}, 2843 (1992).
\bibitem{Chou} C.-H. Chou, T. Yu, and B. L. Hu Phys. Rev. E \textbf{77}, 011112 (2008).
\bibitem{Zerbe} C. Zerbe and P. H\"anggi, Phys. Rev. E \textbf{52}, 1533 (1995).
\bibitem{Dhar} A. Dhar, K. Saito, and P. H\"anggi, Phys. Rev. E \textbf{85}, 011126 (2012).
\bibitem{Wubs} M. Wubs, K. Saito, S. Kohler, P. H\"anggi, and Y. Kayanuma, Phys. Rev. Lett. \textbf{97}, 200404 (2006).
\bibitem{Luczka} J. Luczka, Physica A \textbf{167}, 919 (1990).
\bibitem{vanKampen} N. G. van Kampen, J. Stat. Phys. \textbf{78}, 299 (1995).
\bibitem{Doll} R. Doll, D. Zueco, M. Wubs, S. Kohler, and P. H\"anggi, Chem. Phys. \textbf{347}, 243 (2008); and references therein.
\bibitem{Pauli} W. Pauli, In \emph{Festschrift zum 60. Geburtstage A. Sommerfeld} (Hirzel, Leipzig, 1928).
\bibitem{Lindblad} G. Lindblad, Commun. Math. Phys. \textbf{48}, 119 (1976).
\bibitem{Redfield} A. G. Redfield, IBM J. Res. Dev. \textbf{1}, 19 (1957).
\bibitem{Esposito} M. Esposito and P. Gaspard, Phys. Rev. E \textbf{68}, 066112 (2003).
\bibitem{Blum} K. Blum, \emph{Density Matrix Theory and Applications}, 2nd ed. (Plenum Press, New York, 1996).
\bibitem{Alicki} R. Alicki, and K. Lendi, \emph{Quantum Dynamical Semigroups and Applications}, Lecture Notes in Physics, \textbf{286} (Springer-Publ., Berlin, 1987); see Chapters. II and III therein.
\bibitem{Breuer} H. P. Breuer and F. Petruccione \emph{The Theory of Open Quantum Systems} (Oxford University Press, Oxford, 2002).
\bibitem{Wangsness} R. K. Wangsness and F. Bloch, Phys. Rev. \textbf{89}, 728 (1953).
\bibitem{Laird} B. B. Laird, J. Budimir, and J. L. Skinner, J. Chem. Phys. \textbf{94}, 4391 (1991).
\bibitem{Kohen1} D. Kohen, C. C. Marston, and D. J. Tannor, J. Chem. Phys. \textbf{107}, 5236 (1997).
\bibitem{Fleming1} C. H. Fleming, N. I. Cummings, C. Anastopoulos, and B.L. Hu, J. Phys. A \textbf{45}, 065301 (2012).
\bibitem{Pollard} W. T. Pollard and R. A. Friesner, J. Chem. Phys. \textbf{100}, 5054 (1993).
\bibitem{Kohen2} D. Kohen and D. J. Tannor, J. Chem. Phys. \textbf{103}, 6013 (1995).
\bibitem{Geva1} E. Geva and R. Kosloff, J. Chem. Phys. \textbf{104}, 7681 (1996).
\bibitem{Hanggi2} P. H\"anggi, and G. L. Ingold, CHAOS \textbf{15}, 026105 (2005).
\bibitem{Mori} T. Mori and S. Miyashita, J. Phys. Soc. Jpn. \textbf{77}, 124005 (2008).
\bibitem{Fleming2} C. H. Fleming, and N. I. Cummings, Phys. Rev. E \textbf{83}, 031117 (2011).
\bibitem{Geva2} E. Geva, E. Rosenman, and D. J. Tannor, J. Chem. Phys. \textbf{113}, 1380 (2000).
\bibitem{Romero} V. Romero-Rochin and I. Oppenheim, Physica (Utrecht) \textbf{155}, 52 (1989).
\bibitem{Bogolybov} N. N. Bogolyubov, Publ. Acad. Sci. Ukr. SSR Kiev pp.115–137 (1945), in Russian.
\bibitem{Magalinskii} V. B. Magalinski\u{\i}, Sov. Phys. JETP \textbf{9}, 1381 (1959).
\bibitem{Sentizky} I. R. Senitzky, Phys. Rev. \textbf{119}, 670 (1960).
\bibitem{Ford} G. W. Ford, M. Kac, and P. Mazur, J. Math. Phys. \textbf{6}, 504 (1965).
\bibitem{Zwanzig} R. Zwanzig, J. Stat. Phys. \textbf{9}, 215 (1973).
\bibitem{Caldeira1} A. O. Caldeira and A. J. Leggett, Phys. Rev. Lett. \textbf{46}, 211 (1981).
\bibitem{Caldeira2} A. O. Caldeira and A. J. Leggett, Ann. Phys. \textbf{149}, 374 (1983).
\bibitem{Weiss} U. Weiss, \emph{Quantum Dissipative Systems} (World Scientific, Singapore, 2008).
\bibitem{Jang} S. Jang, J. Cao, and R. J. Silbey, J. Chem. Phys. \textbf{116}, 2705 (2001).
\bibitem{Schroder} M. Schr\"{o}der, M. Schreiber, and U. Kleinekath\"{o}fer, J. Chem. Phys. \textbf{126}, 114102 (2007).
\bibitem{Thingna1} J. Thingna, J.-S. Wang, and P. H\"anggi, (in preparation).
\bibitem{Garcia} J. L. Garc\'{\i}a-Palacios and D. Zueco, J. Phys. A: Math. Gen. \textbf{39}, 13243 (2006).
\bibitem{Campisi} M. Campisi, P. Talkner, and P. H\"anggi, Phys. Rev. Lett. \textbf{102}, 210401 (2009).
\bibitem{Haake} F. Haake, \emph{Quantum Statistics in Optics and Solid State Physics} (Springer-Verlag, Berlin, 1973).
\bibitem{Kubo1} R. Kubo, M. Toda, and N. Hashitsume, \emph{Statistical Physics II - Nonequilibrium Statistical Mecahnics} (Springer-Verlag, Berlin, 1983).
\bibitem{Kubo2} R. Kubo, J. Phys. Soc. Jpn. \textbf{12}, 570 (1957).
\bibitem{Martin} P. C. Martin and J. Schwinger, Phys. Rev. \textbf{115}, 1342 (1959).
\bibitem{Lebowitz} H. Spohn and J. L. Lebowitz, Adv. Chem. Phys. \textbf{38}, 109 (1978).
\bibitem{Kress} R. Kress, \emph{Applied Mathematical Sciences 82: Linear Integral Equations}, 2nd ed. (Springer-Verlag, New York, 1999).
\bibitem{Tanimura2} Y. Tanimura and R. Kubo, J. Phys. Soc. Jpn. \textbf{58}, 101 (1989).
\bibitem{Meier} C. Meier and D. J. Tannor, J. Chem. Phys. \textbf{111}, 3365 (1999).
\bibitem{Honda} D. Honda, H. Nakazato, and M. Yoshida, J. Math. Phys. \textbf{51}, 072107 (2010).
\bibitem{Munro} W. J. Munro and C. W. Gardiner, Phys. Rev. A \textbf{53}, 2633 (1996).
\bibitem{Blanga} L. D. Blanga and M. A. Desp\'{o}sito, Physica A \textbf{227}, 248 (1996).
\bibitem{Davies1} E. B. Davies, Commun. Math. Phys. \textbf{39}, 91 (1974).
\bibitem{Davies2} E. B. Davies, Ann. Inst. Henri Poincar\`{e} \textbf{11}, 265 (1975).
\bibitem{Davies3} E. B. Davies, Math. Ann. \textbf{219}, 147 (1976).
\bibitem{Dumcke} R. D\"umcke, and H. Spohn, Z. Phys. B -- Con. Mat. \textbf{34}, 419 (1979).
\bibitem{Thingna2} J. Thingna (unpublished notes).
\bibitem{Dyson} F. J. Dyson, Phys. Rev. \textbf{85}, 631 (1952).
\bibitem{Peierls} R. Peierls, Z. Physik \textbf{8}, 763 (1933).
\bibitem{Landau} L. D. Landau and E. M. Lifshitz, \emph{Statistical Physics}, 3rd ed. (Pergamon, Oxford, 1980).
\end{thebibliography}
\end{document}